\documentclass[acmsmall,screen]{acmart}
\usepackage{microtype}
\usepackage{hyperref} 

\AtBeginDocument{%
  }
\acmYear{2025}

\begin{document}

\title{Should AI Become an Intergenerational Civil Right?}

\author{Jon Crowcroft}
\authornotemark[1]
\affiliation{%
  \institution{Department of Computer Science and Technology, University of Cambridge}
  \city{Cambridge}
  \country{UK}
}
\email{jon.crowcroft@cl.cam.ac.uk}

\author{Rute C. Sofia}
\affiliation{%
  \institution{fortiss - research institute of the free state of Bavaria for intensive software systems and services}
  \city{Munich}
  \country{Germany}
  }
\email{sofia@fortiss.org}

\author{Dirk Trossen}
\affiliation{%
  \institution{DaPaDOT Tech UG}
  \city{Munich}
  \country{Germany}}
  \email{dirk@dapadot-tech.eu}

\author{Vassilis Tsaoussidis}
\affiliation{%
  \institution{Department of Electrical and Computer Engineering,
  Democritus University of Thrace}
  \city{Xanthi}
  \country{Greece}
}
\email{vassilis.tsaoussidis@gmail.com}

\renewcommand{\shortauthors}{Crowcroft et al.}

\begin{abstract}
Artificial Intelligence (AI) is rapidly becoming a foundational layer of social, economic, and cognitive infrastructure. At the same time, the training and large-scale deployment of AI systems rely on finite and unevenly distributed energy, networking, and computational resources. This tension exposes a largely unexamined problem in current AI governance: while expanding access to AI is essential for social inclusion and equal opportunity, unconstrained growth in AI use risks unsustainable resource consumption, whereas restricting access threatens to entrench inequality and undermine basic rights.

This paper argues that access to AI outputs largely derived from publicly produced knowledge should not be treated solely as a commercial service, but as a fundamental civil interest requiring explicit protection. We show that existing regulatory frameworks largely ignore the coupling between equitable access and resource constraints, leaving critical questions of fairness, sustainability, and long-term societal impact unresolved. To address this gap, we propose recognizing access to AI as an \emph{Intergenerational Civil Right}, establishing a legal and ethical framework that simultaneously safeguards present-day inclusion and the rights of future generations.

Beyond normative analysis, we explore how this principle can be technically realized. Drawing on emerging paradigms in IoT--Edge--Cloud computing, decentralized inference, and energy-aware networking, we outline technological trajectories and a strawman architecture for AI Delivery Networks that support equitable access under strict resource constraints. By framing AI as a shared social infrastructure rather than a discretionary market commodity, this work connects governance principles with concrete system design choices, offering a pathway toward AI deployment that is both socially just and environmentally sustainable.
\end{abstract}

\ccsdesc[500]{Social and professional topics~Computing / technology policy}
\ccsdesc[300]{Social and professional topics~Sustainability}
\ccsdesc[300]{Computing methodologies~Artificial intelligence}
\ccsdesc[100]{Networks~Network architecture}

\keywords{AI, civil rights,
  energy-aware computing, AI governance, networked AI infrastructures}

\received{December 2025}
\received[revised]{2025}
\received[accepted]{2025}

\maketitle

\section{Introduction}

Artificial Intelligence (AI) is no longer a distant technological aspiration. AI is already reshaping healthcare, education, commerce, transportation, economics, and politics. Its rapid evolution constitutes not just a technical advancement but a profound social and economic transformation that touches every dimension of human life.

Ongoing research addresses challenges both internal to AI, including interpretability, bias, robustness, and accountability, and external to AI, affecting scientific domains that are increasingly influenced by deployed AI systems. Representative examples include the dissemination of AI-generated outputs, the architectures and protocols underpinning the networks through which these outputs are transmitted, the integration of real-time sensor data, stringent latency constraints, and the substantial energy consumption of the broader AI ecosystem. There is clear evidence that both  energy and, at least for now, network resources are scarce.

Energy overconsumption has emerged as a central concern (e.g. \cite{IEA2025_EnergyAndAI}), prompting research into networking architectures and systems that reduce AI’s energy footprint. Yet, despite these efforts, the integration of AI into nearly every application domain, the aggressive global expansion of data centers, and the massive investments in AI across all sectors of human activity make it increasingly clear that AI’s cumulative energy requirements will generate a demand that cannot be met without significant societal costs, both for current and future generations.

Within the AI industry, an “easy” response to these pressures is already becoming visible: restricting citizens’ (free) access to AI. Rising usage costs, an increasingly burdensome energy footprint, and escalating demands on network and computational infrastructure, which are already producing bottlenecks, together suggest that access limitations may become inevitable. However, this dimension has been wholly absent from existing regulatory frameworks, including the EU AI Act \cite{EU_AI_Act_2024} and from deliberations within the United Nations \cite{UN2023_GoverningAI} and UNESCO \cite{UNESCO2021_EthicsAI}

However, access to AI output cannot be treated solely as a discretionary market outcome. It arises from a fundamental moral obligation and constitutes a \textit{sine qua non} for the legitimate and equitable use of AI systems. The selective distribution of AI output, or restricted citizen access to critical information, will produce conditions of social inequality, exclusion of vulnerable groups, and discriminatory treatment of individuals. Moreover, if the extraordinary capabilities of AI are made available only to privileged segments of society, this would undermine not only the intellectual rights of citizens, given that AI systems are trained on knowledge they collectively produce and make public, but also their human rights by denying them equal opportunity.

Therefore, it is argued that AI development and citizen access represent two inherently conflicting conditions. On the one hand, rapid AI development without restrictions on access would lead to unsustainable levels of energy consumption. On the other hand, restricting access to AI outputs would infringe on fundamental rights and exacerbate social inequities; a concern that has been largely ignored or at least overlooked.

Hence, this position paper contends that AI development must be explicitly and rigorously bounded by two constraints: (i) citizens’ accessibility and (ii) responsible management of energy resources. The paper argues that this balance cannot be achieved through advisory guidelines or ad-hoc regulatory measures alone; rather, it requires a stable legal foundation. For this reason, it is proposed that access to AI be established as an \textit{Intergenerational Civil Right}. Such a designation would provide a legal and ethical framework that supports equitable policy-making and enables the responsible and sustainable development of future AI technologies.

The main contributions of this paper are:
\begin{itemize}
    \item A normative claim: access to AI should be recognized as an Intergenerational Civil Right.
    \item A governance insight: existing AI regulation systematically ignores energy, networking, and access coupling.
    \item   A systems contribution: a decentralized, energy-aware AI Delivery Network (AIDN) architecture aligned with public-infrastructure principles.
\end{itemize}

The paper is organized as follows. Section~\ref{citizenaccess} examines why citizen access, although widely taken for granted today, will inevitably become progressively constrained as demand increases and available resources become insufficient. We focus specifically on energy consumption and network congestion, which, in our view, will lead to access limitations. Section~\ref{regulation-gaps} discusses the implications of selective AI access and highlights gaps in the current regulatory landscape, particularly its shortcomings regarding energy and networking resources. Section~\ref{AI-Right} elaborates on the concept of an Intergenerational Civil Right and explains how it could function as a balancing mechanism between these two boundaries. Section~\ref{sec:trajectories} presents candidate technological trajectories that support AI as a shared social infrastructure under such a framework, emphasizing emerging paradigms in networking and computing. Section~\ref{sec:strawman} introduces a 3D architecture and a strawman design for AI Delivery Networks (AIDNs) that operationalize these principles at scale. Finally, Section~\ref{conclusions} summarizes the position advanced in this paper and provides key takeaways.
 
\section{Why Citizen Access Will Shrink Over Time}
\label{citizenaccess}
\subsection{Scaling of AI Demand}

Much has been written about the scaling limits for AI training (e.g. \cite {Kaplan2020ScalingLaws}) but let us first focus our discussion on the growing demand for AI inferencing (over existing and evolving AI models) over time. 

For this, we conduct a simple \textit{Gedankenexperiment} that looks at a possible proliferation of AI inferencing functionality being integrated into mobile applications that we use everyday. More specifically, AI inference is foreseen to become embedded in a wide range of functions currently delivered through mobile applications. A similar evolution has occurred with mobile mapping, which has expanded well beyond stand-alone applications such as Google Maps and is now integrated into transportation, travel, fitness, and logging apps, among many others. This integration has substantially increased the demand for cartographic capabilities and services.

We start with the assumption of a mobile user through its mobile applications to generate 60 requests per second at peak rate, thus 3600 RPM (requests per minute). 

Let us now assume that roughly 30\% of smartphones will enable AI capabilities. With a smartphone population of about 70\% out of 7 billion end user subscriptions in 2025, up to 1.47 billion active AI users may initiate AI requests. Joining this with our expected peak usage rates, mobile AI inferencing alone may generate up to 5.292 trillion requests per minute (or 88 billion requests per second), if we assume wide-spread integration into mobile application functionality. 

Turning from the generation of inferencing requests back to the AI training problem, localized data, such as from Internet of Things (IoT) cyber-physical systems, smarthomes, energy grids, transportation systems, cars, and many more digitally connected objects, will equally drive not just training itself but the ingestion traffic generated from those data sources into training models. Here, we foresee not just the ingestion into centralized models, much like how today's GPT-like systems work, but the fine-tuning of already (often centrally) trained models through locally available and relevant information. Take the example of traffic congestion management in road systems, where training models for large scale prediction may help planning objectives, while locally fine-tuned models, utilizing congestion data from moving cars, may be used for local traffic management. Through this, an AI ecosystem will not just observe growing inferencing but also a constant stream of ingestion data from distributed resources towards locations where suitable AI models are trained and adjusted.    

\subsection{Communication Bottlenecks}

The scaling of AI demand, as discussed in the previous subsection, introduces multiple bottlenecks driven by rising traffic associated with both AI inference and data ingestion.

These bottlenecks are driven by the current centralization of capabilities, particularly in large-scale data centers. If provided through only few locations, our observed rising demand for AI traffic will create very large incast traffic into relatively few inferencing Points of Presence (PoPs), which in turn will lead to congestion and increase in latency, therefore impacting the token response time which is an important AI system performance metric. 

Upgrading the communication infrastructure to cope with those foreseeable incast and congestion problems is itself a bottleneck in regards to coping with the exploding AI traffic demand. The growth rates of AI services, such as observed in users for chatGPT\footnote{https://www.tooltester.com/en/blog/chatgpt-statistics/}, are exceeding those of early social media sites in the mid 2000s. Keeping up with this demand at the level of the communication infrastructure is expensive, despite advances from the current, e.g., mobile 5G infrastructure to the future 6G systems currently being standardized.

From both technological and economic perspectives, significant challenges arise at the infrastructure level. Current pricing models largely focus on the volume of data consumed and the bandwidth over which it is delivered. In the mobile market, pricing is typically based on data caps measured in gigabytes, while residential broadband services are commonly constrained by bandwidth limits.

In contrast, AI services are primarily driven by latency requirements, with data volume becoming a concern mainly when rich media is involved, such as generative AI–based video or image content. Supporting high-performance AI services therefore requires a shift in emphasis toward fast and efficient delivery. This shift must also align with users’ expectations for energy efficiency, which leads directly to the issue explained in the next sub-section.

\subsection{Energy Requirements of AI Ecosystems}

Similar to AI training, inference is a costly process from an energy
perspective. Prior estimates suggest that an AI-based search request may
consume up to three orders of magnitude more energy than a conventional
search query\cite{AIStackExchangeEnergy}.

At the same time, inference latency is typically on the order of a few
milliseconds per generated token, with the response performance of GPT-class
systems commonly expressed in tokens per second (TPS).
Certainly, efforts to develop more efficient training and inference methods, such as request batching and the use of context caches, for example in the form of key–value (KV) caches, will help curb the growing demand associated with AI services.

A primary driver of energy consumption in today’s AI ecosystems is the high degree of centralization in  AI training and, increasingly, inference over trained models. Although the \textit{Landauer principle}\footnote{\url{https://en.wikipedia.org/wiki/Landauer\%27s\_principle}} establishes a theoretical lower bound on the energy required for computation, including AI, practical factors play a far more significant role. These include the geographic location of training data centers, local energy prices, and regulatory and policy environments.

Recent developments illustrate this dependence. For example, renewed interest in nuclear energy such as the reinstatement of the "Three Mile Island" nuclear plant to support Microsoft’s growing AI workloads, may slow down energy scarcity for large-scale data centers in certain regions. However, such approaches are not universally applicable, particularly in countries where nuclear power is unavailable or has been phased out, as in Germany’s complete nuclear shutdown in 2023\footnote{Germany, as a key economic power, including for the development of AI technologies, decommissioned nuclear energy in 2023: \url{https://www.base.bund.de/en/nuclear-safety/nuclear-phase-out/nuclear-phase-out\_content.html}}.

At the same time, nuclear fusion, long a subject of sustained research and development, is experiencing renewed momentum, in part driven by advances in AI technologies themselves\footnote{\url{https://www.iter.org/node/20687/ai-ignites-innovation-fusion}}. While promising, fusion remains a longer-term prospect and does not resolve near-term energy constraints for AI infrastructure.

Thus, we can see right now an alignment of energy production to the key architectural concepts underlying today's AI systems, namely that of strong centralization, both of energy production and computation.  

\subsection{Mechanisms through Which Access Will Be Restricted}

Controlling demand to match the supply is a well-known techno-economic problem. Methods like pricing for tiered access, as well as subscription models (to rate limit access), and geographic division of the market to locally steer and limit traffic are well-known mechanisms. 

Further, institutional affiliation may play a role, be it through the use of traffic engineering capabilities that is limited to certain organizations\footnote{The authors observed that universities, for instance, use explicitly set routes to GPT-based AI services for improved performance compared to a private user access.} or through institutional level participation in a preferential pricing scheme. 

While balancing demand and supply is key to establish a working market (for knowledge), we argue in the remainder of this paper that a balance between commercial and societal benefits needs to be achieved to ensure continued and fair access to the growing knowledge captured in AI systems. 

\section{Gaps in Current Regulation: The Moral Gap and Social Risks of Selective AI Access} 
\label{regulation-gaps}
Although current AI policy frameworks address issues such as safety, accountability, risk, and energy management, they remain largely silent on the question of equitable access. This omission creates a significant moral gap. AI systems are becoming essential for education, employment, healthcare, scientific discovery, and civic participation, yet there is no regulatory assurance that these capabilities will remain accessible to all. We argue that selective restriction of access to AI lacks moral justification, poses substantial social risks, and remains insufficiently addressed by existing national and international regulatory frameworks.

\subsection{Social and Ethical Implications of Selective Access}
Restricting access to AI systems introduces new forms of digital exclusion (e.g. \cite{Eubanks2018AutomatingInequality}). As AI becomes embedded in everyday applications, from job recruitment systems to educational platforms, public services, and decision-making tools, unequal access directly translates into unequal opportunities. Populations already facing structural disadvantages, such as low-income communities, marginalized groups, and individuals with limited digital literacy, are at increased risk of falling further behind. Yet, the risks extend beyond traditionally vulnerable populations. Even highly educated individuals, who have invested significant effort in acquiring expertise, may experience a form of meritocratic erosion: the comparative advantages they previously held can be neutralized or even reversed, and conventional assessments of capability and skill may be displaced by access-driven performance.

Selective access therefore generates not only profound knowledge divides, but also a systematic undermining of merit. Individuals able to afford premium AI services effectively “borrow” enhanced analytical, educational, and creative capabilities, often surpassing the abilities of those who intrinsically possess such skills or have cultivated them through years of study. The outcome is an asymmetry of informational power that unjustly entails privilege. Knowledge that should function as a public good is transformed into a scarce, purchasable resource, subverting societal commitments to equal opportunity and distorting the very foundations of meritocracy.

\subsection{Moral Obligations Regarding Public Knowledge}
A central moral tension arises from the fact that AI systems are predominantly trained in publicly produced knowledge \cite{InformationDemocracy2024} This includes texts, images, research outputs, cultural artifacts, and information that citizens have generated, published, or contributed to over decades and centuries. The public corpus is the substrate on which modern AI is built; without it, large-scale models could not exist (\cite{WeylLanier_PublicOptionAI}.

Moreover, the intellectual property system itself acknowledges that knowledge ultimately belongs to the public: patents and copyrights expire, returning scientific and cultural works to the public domain. This principle reflects the idea that innovation is cumulative and that society has a right to access foundational knowledge (e.g. \cite{IamIP2025WhyPatentsExpire} and \cite{Price2017ExpiredPatents}).
Public libraries represent the most familiar institutional expression of this idea. They ensure that knowledge—once protected works enter the public domain—is available to all, regardless of socioeconomic status. By analogy, AI systems trained on publicly accessible knowledge should not become tools available only to those with the economic means to purchase access. Restricting AI outputs is to privatize the benefits of publicly created knowledge, violating both the spirit of intellectual property law and long-standing democratic norms regarding access to information.

\subsection{Regulatory Gaps and Limitations}
Existing regulations primarily focus on preventing discrimination, unsafe deployment, or lack of transparency; however, they bypass the critical ethical question of whether citizens have a justifiable claim to access AI systems derived from collectively produced public knowledge. In particular, no regulatory instruments acknowledge the intellectual contributions of the public, and there is no mandate to ensure that AI systems trained on public data provide proportionate public benefit.

More specifically, the EU AI Act (Regulation (EU) 2024/1689) is structured around a risk-based approach: it classifies AI systems into categories (unacceptable, high risk, limited risk, minimal risk) and imposes different obligations accordingly. However, the Act does not guarantee a \textit{“right of access}” for citizens to AI services. It places duties on providers (e.g., high-risk systems must comply with transparency, safety, and quality requirements) but does not mandate universal or affordable provision of AI capabilities.
The United States does not yet have a single, comprehensive federal law governing AI systems, resulting in a regulatory environment fundamentally different from the EU’s harmonised approach. Instead, the U.S.A AI governance emerges from a combination of voluntary standards, most notably the NIST AI Risk Management Framework \cite{NIST_AIRMF2023}, executive-branch directives, such as the White House’s Executive Order on the Safe, Secure, and Trustworthy Development and Use of Artificial Intelligence \cite{USEO2023}, and sector-specific regulation enforced by agencies like the FTC, FDA, and others operating under pre-existing statutory authority \cite{CRS_USAI}. In addition to federal action, numerous state-level initiatives have introduced a patchwork of rules governing the development and deployment of AI, further contributing to regulatory fragmentation between jurisdictions \cite{State_AI_Overview}. This decentralized model provides flexibility and aims to facilitate innovation, however, it produces uncertainty regarding compliance obligations and overlooks civil rights.

UNESCO’s Recommendation on the Ethics of Artificial Intelligence \cite{UNESCO2021_EthicsAI} emphasizes human rights, fairness, and social justice, and explicitly calls for “equitable access” in its policy action areas. It encourages Member States to foster multi-stakeholder inclusion, mitigate digital divides, and make AI benefits accessible to marginalized and vulnerable groups. However, the Recommendation is voluntary, not legally binding: UNESCO encourages states to take “legislative or other measures … as required”, but does not impose enforceable rights to AI access. Although it stresses the importance of safeguarding the interests of present and future generations, it lacks concrete mechanisms to operationalize intergenerational fairness. In particular, it does not define quantitative, technical, or legal instruments to limit energy consumption, manage resource use, or ensure that such constraints do not diminish equitable access over time. Recent works (e.g. \cite {Freilich2025_DataAsPolicy}) highlight the need to shift to new policy paradigms, such as, for example, to utilize data itself as a regulatory instrument to implement policies that guarantee social benefits.The authors in \cite{ Vinuesa2019_AIforSDGs} make an important contribution towards both sustainability and equitable access: they evaluate how AI affects each of the UN Sustainable Development Goals (SDGs), concluding that AI has significant potential to produce global public-good benefits, such as improving education, healthcare, climate modeling, and resource management. However, they warn that AI can also undermine sustainability if governance, ethical safeguards, and equitable access are not ensured.

Furthermore, current regulatory frameworks also fail to explicitly account for upcoming bottlenecks in AI inference distribution arising from energy, network, and computational constraints. By leaving the administration of AI-related resources largely to the discretion of private owners, these frameworks risk transforming AI, a powerful tool built upon public knowledge, into a privately controlled resource. Without mechanisms to ensure equitable access, unsustainable energy use,  and intergenerational responsibility, the existing regulatory landscape leaves ample space to violate civil rights, human rights, and the rights of future generations.

\subsection{The Flawed Assumption of Universal Accessibility}
A pervasive, but largely unexamined assumption in the current AI debate is that access to AI systems will become universal by default~\cite{vspecian2025universal}. This assumption underlies much of the optimism surrounding the potential of AI-driven applications to support and augment education, expand economic opportunities, and democratize knowledge. However, evidence starts to show some barriers that need to be addressed as quickly as possible. Beyond the energy barrier and computational resource availability, structural barriers already limit who can use, access, or fully benefit from AI technologies. These barriers reveal that universal accessibility is not simply threatened by future energy or network constraints; it is already compromised today. They also underscore why regulatory and ethical frameworks cannot rely on market forces or technological progress to guarantee equitable access.

One of the most pressing gaps concerns the skills and infrastructural readiness of educational systems around the world~\cite{bo2025oecd}. Schools and universities are increasingly expected to integrate AI into learning environments. However, as corroborated by initiatives such as the ACM "Informatics for AlL"\footnote{https://europe.acm.org/i4all} a large number of schools and universities, if not the majority, lack the pedagogical frameworks, teacher training, and technical infrastructure necessary to support an effective application and use of informatics, situation which is even more dramatic with AI. These disparities mirror broader socioeconomic inequalities: while affluent institutions can leverage AI to enhance learning, under-resourced schools risk falling further behind. Instead of functioning as an equalizer, AI risks amplifying existing gaps in digital literacy and educational opportunity. If future access becomes restricted or tiered, these inequalities will deepen dramatically, creating generations whose cognitive and professional development is constrained by their institutions’ inability to incorporate AI.

Language and accessibility barriers further undermine the notion of universal access. Despite rapid progress, AI systems still disproportionately serve speakers of high-resource languages, especially English~\cite{languages}. Users who speak regional or minority languages, or who require accessibility features such as simplified interfaces, assistive technologies, or multimodal support, remain systematically disadvantaged. This linguistic imbalance effectively allocates AI’s benefits along geo-political lines: linguistic and cultural communities that are underrepresented in training  may end up receiving lower-quality outputs, reduced reliability, and diminished utility. Hence, even before any formal access restrictions emerge, millions of individuals may already experience a de facto form of limited access due to language accessibility shortcomings.

A third factor challenging universal accessibility is the increasing \textit{commodification of AI}. As models become more capable and resource-intensive, providers are shifting toward subscription-based access, usage quotas, pay-walled features, and tiered service levels. This business model treats AI as a commercial product rather than a public good, creating stratified access regimes that distribute cognitive and informational advantages according to purchasing power. Premium versions of AI systems offer greater accuracy, reliability, speed, or reasoning capability, and will grant economically privileged users capabilities that others simply cannot afford. In effect, knowledge and problem-solving ability risk becoming market commodities, purchasable only by those with sufficient resources.

Finally, universal accessibility is undermined by the massive infrastructural and energy requirements of AI systems. Even if economic and linguistic barriers were addressed, AI access would remain contingent upon extensive computational infrastructures, namely, hyper data and distributed cloud centers that have resilient energy supplies, that are unevenly distributed across regions and countries, and that may not adapt well to the specific economical and societal fabric of different continents, as occurs in Europe. As demand continues to grow, these infrastructures will become increasingly strained. Regions with limited energy capacity, unstable network infrastructure, or insufficient investment will face increasing access limitations. Access depends on vast and resource-intensive technological ecosystems that cannot be scaled infinitely or equitably without careful governance and planning.

Recognizing these limitations is essential for understanding why equitable access cannot be left to market dynamics or short-term policy interventions. It strengthens the argument for an intergenerational civil right, which alone can provide a durable framework to ensure that access to AI remains equitable, sustainable, and socially just across present and future generations.
 
\section{AI as an Intergenerational Civil Right}
\label{AI-Right}
\subsection{AI as Both a Right and a Threat}
AI technologies now mediate core social functions, including education, employment, administrative services, political participation, and access to information. As a result, the ability to use AI systems increasingly serves as a condition for equal opportunity, similar to access to digital communication, education, or public knowledge. In this sense, AI access embodies an emerging civil right: without it, individuals risk being excluded from essential social, economic, and political processes \cite{UN_UDHR_1948}. Because AI models are trained on collective human knowledge, including publicly funded research and cultural data, citizens have a legitimate normative claim that the benefits of AI should be broadly and equitably distributed.

Yet, this civil-rights dimension coexists with a profound environmental and intergenerational threat. The energy consumption associated with training and deployment of large-scale AI systems is accelerating at a pace that places significant pressure on power grids, renewable energy capacity, and carbon-emissions targets. If left unchecked, widespread and unconstrained AI proliferation could accelerate energy depletion and ecological harm to such an extent that the basic rights and living conditions of future generations are undermined. In this sense, unregulated AI access may violate intergenerational justice.
\subsection{Conceptual Definition}
Current regulatory structures are not equipped to manage this duality. AI governance instruments typically focus on risk mitigation, discrimination prevention, and transparency obligations, but do not integrate civil-rights protections with environmental safeguards. As a result, a regulatory gap emerges: when AI is treated solely as a civil right, universal access can encourage unsustainable energy demand; when treated solely as an environmental threat, restrictive measures can entrench inequality and allow private actors to monopolize AI capabilities.

We propose that access to AI be recognized as an intergenerational civil right, serving a dual protective function. First, it would ensure that AI systems are deployed in ways that avoid excessive energy consumption that could endanger the sustainability of the planet and thus compromise the well-being of future generations. Second, it would provide a legal framework for equitable access for all citizens today, preventing AI from becoming a resource monopolized by a privileged minority. In this sense, an intergenerational civil right integrates the responsibility to preserve environmental and energy resources with the civil imperative to maintain equal opportunity.

\subsection{The Balancing Mechanism}
The Intergenerational Civil Right (ICR) could serve as a balancing mechanism to reconcile AI’s dual nature: as a civil right and as a threat due to its energy-intensive footprint.

First, the ICR can establish fairness models for distributing AI inferences and outputs among users. By embedding principles such as proportional, max-min, or weighted fairness into allocation systems, it ensures that access is not monopolized by privileged actors and that all citizens benefit equitably from AI capabilities.

Second, the ICR provides a framework for monitoring and controlling energy consumption. It enables granular measurement of AI energy usage, sets thresholds for responsible deployment, and guides policies to prevent excessive consumption, whether by individual actors or in aggregate.

Third, the ICR serves as a research and investment directive. By signaling the dual imperatives of fairness and sustainability, it incentivizes the development of energy-efficient AI architectures, low-energy inference techniques, and public-interest AI models, shaping technological and market trajectories toward socially responsible innovation.

Finally, the ICR has indirect social and economic effects, particularly in the labor domain. Recognizing AI as a civil right influences policy decisions around automation, including mechanisms such as robot taxes or contributions to social care funds, ensuring that the societal benefits of AI are equitably shared while mitigating negative repercussions on employment and social equity.

In summary, the ICR provides a holistic framework that ensures equitable access while mitigating AI’s associated risks.

\subsection{AI as a Shared Social Infrastructure}
A fundamental premise of the current proposal is that \textit{AI must be understood not merely as a technological product or a commercial service but as a form of shared social infrastructure}. This reframing is essential for two reasons. Firstly, AI increasingly mediates participation in central domains of the societal day-to-day, e.g., knowledge exchange and acquisition, communication, employment, education, public administration, or even in access to essential services. Secondly, AI systems are built upon collective input: public data, publicly funded research, communal linguistic and cultural artifacts, and the cumulative intellectual contributions of society. AI is therefore starting to be a pillar of public knowledge infrastructures such as public libraries, broadcasting, and national education systems and, as a consequence, of human cognitive evolution \cite{riva2025invisible}.

Social infrastructures are characterized by three properties: (i) support fundamental social functions; (ii) rely on collective resources; and (iii) generate obligations of equitable provision. AI systems increasingly meet all three conditions. FAI has become a prerequisite for effective participation in contemporary society. Citizens without access to AI-supported learning tools, job-matching systems, health information platforms, and automated administrative services will face growing disadvantages compared to those that can fully use and understand AI tools. Then, AI development depends directly on public resources: AI factories, datasets that come from citizen-generated content, research conducted in publicly funded institutions, and linguistic and cultural knowledge encoded in shared digital spaces. Therefore, AI services are built on individually and collectively produced inputs: the expectation is that the outputs should not be monopolized by private actors or restricted to privileged groups.

Understanding AI as infrastructure also highlights an essential tension: unlike traditional infrastructures such as libraries or public schools, AI relies on globally distributed computational and energy-intensive systems with substantial operational costs. This dual nature — as both a public good and a resource-intensive technology — means that AI cannot simply be integrated into existing infrastructure frameworks without new governance mechanisms. 

Importantly, conceptualizing AI as shared social infrastructure also aligns with broader historical patterns of knowledge as public goods \cite{stiglitz1999knowledge}. Knowledge infrastructures have traditionally emerged when the social value of broad-based access outweighs the benefits of exclusive control. Public libraries were created to democratize access to printed knowledge; public broadcasting systems were established to ensure equitable information access; universal education became a cornerstone of democratic participation. Today, AI occupies a similar position: its capabilities are too central to civic life, social mobility, and human development to be treated as an optional service or as a luxury commodity.

Recognizing AI as shared social infrastructure  reinforces the moral and political claim that all citizens should have access to the capabilities derived from systems trained on collective knowledge. It also strengthens the argument that such access must be safeguarded by stable, enforceable rights, in particular due to the fact that resource constraints threaten to create new forms of exclusion. In this sense, AI as infrastructure is not merely a descriptive category but a normative imperative: it compels us to design governance systems that balance social equity with ecological responsibility, ensuring that AI remains a shared resource accessible to both present and future generations.

\section{Technological Trajectories to Support AI as an Intergenerational Right and Public Social Infrastructure }\label{sec:trajectories}

Recognizing access to AI as an intergenerational civil right implies that AI must be treated not as a commercial product but as a form of public social infrastructure. This reclassification fundamentally alters the way AI technologies must evolve. Technological development can no longer be driven exclusively by performance benchmarks, proprietary control, or market demand; instead, it must be oriented toward two societal priorities: (i) guaranteeing equitable access for all citizens and (ii) ensuring responsible, sustainable use of the underlying data, networking, and computational infrastructures upon which AI depends.

This shift demands a structural transformation of the AI ecosystem. If access to AI is a protected civil right, then AI systems must expand to satisfy the same design constraints as other rights-bearing infrastructures. The AI systems available today are based on large monolithic models executed in central cloud data centers and are neither energy sustainable nor socially equitable. They depend on high-energy GPU clusters, intensive data pipelines, global network bandwidth, and repeated cycles of model retraining. Under these conditions, access will inevitably shrink over time, unless the architecture, deployment model, and orchestration of AI systems are redesigned from first principles.

The HiPEAC Vision 2025 \footnote{https://www.hipeac.net/vision/2025/introducing-hipeacs-vision-for-the-future-the-next-computing-paradigm/\#/}  articulates the contours of this shift through its proposal for a \textit{Next Computing Paradigm (NCP)}: a computing continuum that seamlessly integrates IoT devices, edge computing resources, cloud infrastructures, and AI accelerators into a federated and sustainable whole. This continuum is not merely a technical convenience; it is the only viable foundation to support AI as a public utility. In this model, AI workloads become mobile, modular, and context-aware, dynamically migrating to the most appropriate layer of the continuum depending on latency, energy availability, fairness requirements, and user needs.

\subsection{Decomposed AI Workloads and Processes for Public Orchestration}

To function as a social infrastructure, AI must become orchestratable. Today’s AI systems execute inference as a single, opaque process inside large models hosted in distant cloud servers. This monolithic design is incompatible with equitable access: it centralizes control, concentrates energy consumption, and creates single points of failure. Instead, a public AI ecosystem must decompose inference, reasoning, and retrieval into smaller computational units that can be executed throughout the computing continuum.

Under a decomposed approach, lightweight AI components, such as embedding, tokenization, shallow attention, simple routing, will run at the edge, near users, and communities. More specialized modules, such as retrieval layers or planning engines, run in regional micro-data centers. Only the most compute-intensive reasoning operations execute in cloud-scale environments, and even then subject to energy budgeting, scheduling, and fairness constraints. This structure mirrors the evolution from monolithic applications to the microservices architectures and orchestration solutions that now underpin the so-called IoT-Edge-Cloud continuum, enabling scalability, transparency, and resilience \cite{sofia2024framework}.

Recognizing AI as a social infrastructure also requires rethinking how the computational processes of AI can be better orchestrated: this includes considering the potential decomposition of inference, reasoning, and retrieval. In the current paradigm, most AI systems rely on large monolithic models that execute inference as a single, opaque process, often on distant cloud servers. However, if AI is to function as a sustainable and equitable public infrastructure, the internal structure of inference itself must be decomposed and orchestrated more intelligently. Drawing on principles from distributed systems and the next computing paradigm, AI computations can be partitioned into smaller, modular components that run on different layers of a heterogeneous computing continuum: lightweight components at the edge, specialized reasoning modules in regional micro-data centers, and high-complexity tasks reserved for energy-intensive cloud environments only when strictly necessary. 

This decomposition mirrors the evolution from monolithic applications to microservices architectures, which currentlz start to enable scalability, transparency, modularity, and resilience across the IoT-Edge-Cloud continuum. Projects such as the Horiyon Europe Context-aware Decentralized Edge-Cloud Orchestration (CODECO) project \cite{sofia2024framework}\footnote{https://gitlab.eclipse.org/eclipse-research-labs/codeco-project} alreadz demonstrate how modular workloads can be orchestrated across heterogeneous infrastructures using real-time telemetry, energy-awareness, and contextual signals. Decomposition is therefore not only a technical decision but a governance mechanism: it distributes control, mitigates single points of failure, and ensures that AI workloads can adapt to the needs and constraints of different communities.

A similar shift is required in the realm of reasoning. Rather than executing all reasoning steps within a single large model, reasoning can be decomposed into sequential or parallel modules, each optimized for a specific cognitive function: retrieval, deduction, planning, verification, or multimodal grounding. Compositional reasoning pipelines, where simpler models handle straightforward tasks, and more sophisticated or energy-intensive modules are invoked only when needed, yield substantial reductions in energy consumption while preserving or even improving performance. 

In the context of AI as a social infrastructure, decomposed reasoning enables fairer allocation of computational resources, supports transparent governance, and aligns model execution with the constraints of an intergenerational civil rights framework. It ensures that the “intelligence layer” of society is not concentrated in a small number of opaque, high-consumption systems but is instead implemented through orchestrated, modular, energy-conscious components accessible to all.

\subsection{KV Caches as the State Mechanism for Distributed Inference}
A crucial enabler of decomposed and energy-efficient inference is the growing use of \textit{key–value (KV)} caches, which store intermediate representations, namely, attention keys and values produced during earlier inference steps. In conventional transformer-based models, each new token requires recomputing attention scores against all previous tokens, generating computational and energy costs that grow quadratically with sequence length. Techniques such as FlashAttention \cite{dao2022flashattention} have shown that a substantial fraction of inference costs stem not from arithmetic computation but from repeated memory operations. KV caches eliminate much of this redundancy: by persisting intermediate attention states, models can process new inputs with only incremental computational overhead, dramatically reducing energy use and latency.

Recent work demonstrates that KV caching is not only a performance optimization but a central mechanism for scaling large language models efficiently. Studies on KV cache compression and quantization \cite{shi2024keep} demonstrate that caches can be significantly reduced in precision or size without meaningful loss in output quality. Long-context optimizations \cite{an2024make} further show that effective cache management is essential to make xtended reasoning sequences feasible under realistic energy constraints.

Within an AI social-infrastructure framework, KV caches provide a mechanism for portable cognitive state. Cached attention states can be transmitted, offloaded, or resumed across heterogeneous Edge-Cloud environments. Hence, KV-based AI architectures enable inference to be decomposed across a computing continuum in line with the emerging NCP proposals. Early-stage inference tasks—tokenization, embedding, shallow attention can be executed at the edge, while deeper layers or more complex reasoning modules run only when needed on higher-capability hardware. This aligns with insights from memory-augmented and recurrent alternatives to transformers, such as RWKV \cite{peng2023rwkv} and Memorizing Transformers \cite{wu2022memorizing} , which show how contextual state can be externalized and reused.

\subsection{Sustainable AI Architectures}

If AI is to operate as a public social infrastructure, its underlying model architectures must be engineered to satisfy requirements that differ fundamentally from those of commercial, performance-maximizing systems. Instead of optimizing for peak benchmark accuracy or usage scale, sustainable AI architectures must support universal access under strict resource constraints. This implies designing models that are energy-aware, hardware-adaptive, composable across the computing continuum, and maintainable by public institutions lacking hyperscale computing capacity. Prior work has shown that large model training and inference already produce substantial energy costs and environmental burdens \cite{strubell2019energy, patterson2021carbon} underscoring that sustainability is not optional but a structural requirement for the long-term provision of AI as a public good.
Sustainable AI architectures must therefore be optimized not only for accuracy but for resource fairness, energy sensitivity, locality of execution, and capability to distribute workloads based on context-awareness and user requirements across heterogeneous compute environments.

A first requirement is deployability on heterogeneous and resource-limited environments. Public institutions such as schools, municipalities, hospitals, libraries often operate with constrained computational budgets and uneven connectivity. Techniques such as model compression, pruning, quantization, distillation, and tensor factorization have been shown to dramatically reduce model size and inference cost \cite{han2015deep}. Parameter-efficient fine-tuning approaches, including adapters and LoRA-style methods \cite{hu2022lora} are a way to enable institutions to adapt foundation models without expensive retraining. 

A second requirement is architectural modularity. Monolithic models are difficult to audit, update, or govern, and their retraining cycles impose substantial and recurring energy costs. Modular architectures incorporating components for retrieval, domain adaptation, multi-modal processing, or reasoning enable targeted upgrades without full retraining. This mirrors emerging research on modular retrieval-augmented generation \cite{gao2024modular}, mixture-of-experts models \cite{mu2025comprehensive}, all of which demonstrate that modularity can improve efficiency while supporting flexible, compositional reasoning.

Third, sustainable AI architectures must adopt tiered computational and networking pathways, where lightweight models or shallow inference routes handle routine requests, reserving deeper or more energy-intensive layers for complex tasks. Recent work on early-exit architectures, adaptive computation, and selective activation \cite{rahmath2024early} shows that significant energy savings can be achieved without substantial loss in accuracy. Simple requests do not monopolize compute, and energy-intensive reasoning is triggered only when justified by social or task-specific value.

Fourth, sustainable AI depends on architecture that enables orchestration across the computing continuum—a priority emphasized in the HiPEAC NCP vision and central to reference IoT-Edge-Cloud architectures such as the one embodied in the CODECO illustrates how AI workloads can be decomposed and scheduled dynamically across heterogeneous nodes, from IoT devices, considering edge servers, regional micro-data centers, and cloud clusters based on contextual factors such as energy availability, latency requirements, network congestion, or user location. This decentralized orchestration is essential for public AI infrastructure: it ensures that computation occurs close to users when possible, reduces long-distance data movement (a dominant source of energy consumption), increases resilience, and prevents the monopolization of compute resources by centralized providers. Context-efficient inference, including intelligent handling of long contexts, selective state retention, and low-footprint KV cache management plays therefore a key role in this context. 

A fifth requirement is hardware-awareness and co-design. Research on heterogeneous computing such as NCP demonstrates that aligning model architectures with the characteristics of target hardware—edge accelerators, NPUs, low-power embedded processors, analog or neuromorphic chips can greatly reduces energy consumption by minimizing data movement, a dominant source of energy cost. NCP proposed by the HiPEAC Vision emphasizes that future computational systems must operate along the IoT-Edge-Cloud continuum  and hence, specialized accelerators with workloads dynamically assigned to the most appropriate hardware based on locality, energy availability, and latency requirements. 

Finally, sustainability requires lifecycle governance, an often overlooked but essential architectural constraint. Unlike commercial systems built on short release cycles, public AI utilities must commit to long-term stability, predictable environmental footprints, and backward compatibility. This requires explicit guidelines for model updating, archival, expiration, version control, and environmental accounting \cite{patterson2021carbon}. Architectural choices therefore must support maintainability: modular retraining, quantization-aware designs, and mechanisms that allow institutions to ensure that models continue to satisfy fairness and energy requirements across their entire operational life, in a way that is common to any other energy consuming goods provided to the public citizen.

\subsection{Networking and Data Infrastructure for Public AI}
For AI to function as a public social infrastructure, the underlying networking and data infrastructure must be re-thought as foundational public assets rather than commercial substrates. The current cloud-centric arrangement, which depends heavily on long-distance data transmission and centralized inference endpoints, is already showing signs of structural strain. Studies consistently demonstrate that data movement, rather than computation, is one of the dominant contributors to energy consumption in modern AI workloads \cite{10630483}. As the scale of AI use expands, reliance on central servers heightens latency disparities, produces network congestion, and amplifies inequalities between well-connected regions and communities with limited bandwidth or infrastructure. A public AI ecosystem must therefore embed intelligence not only within models but within the network itself.

The NCP paradigm provides a conceptual roadmap for this transformation. In the NCP, computation becomes mobile: workloads migrate fluidly across IoT devices, local edge nodes, regional micro–data centers, and edge-cloud platforms. Networking is no longer a passive conduit but an active decision-making layer that determines where data should be processed, how inference should be routed, and which resources should be activated under specific energy, latency, or fairness constraints. To support AI as a civil right, the network must make these decisions in ways that minimize long-distance traffic, prioritize local execution when feasible, and ensure that all users, regardless of their location and means, can receive reliable and timely AI responses.

Furthermore, it is important that the overall infrastructure architecture can support, in an integrated way, the distribution and orchestration of AI workloads and processes. Related literature, for instance, provides work on model and output caching demonstrates that shared inference results can reduce compute load by an order of magnitude for common tasks \cite{li2024llm}. 

Public Retrieval-Augmented Generation (RAG) architectures rely on large vector databases and indexing layers, which are currently operated primarily by private cloud providers \cite{oche2025systematic}.  For AI to operate as a social infrastructure, these retrieval layers must become public utilities themselves, accessible to educational institutions, municipalities, and civic organizations. Public vector databases, e.g., to be replicated or made available regionally, energy-budgeted, and governed transparently can ensure that communities can access high-quality retrieval without submitting all queries to centralized corporate servers. 

Notions such as in-network caching and forwarding are therefore relevant in this context. Caching will be required across different networking layers to reduce downloads, avoid redundant inference, and long-distance routing. Caching in this context needs to be extended beyond model caching, and as explained, KV caches are essential in this context.

This layered caching resembles the logic of Content Delivery Networks (CDNs), but optimized for generative and retrieval-based AI. In the public-infrastructure context, the recent concept of AI Inference Defined Networks (AIDNs) is a must for which we outline a first strawman architecture in Section \ref{sec:strawman}. Analogous to CDNs but optimized for generative and retrieval-based AI workloads, AIDNs introduce caching at multiple layers of the infrastructure. Model weights, vector embeddings, generated outputs, and even KV caches can be stored in regional nodes, significantly reducing redundant computation and avoiding repeated traversals of long-haul links. Research on inference caching shows that these mechanisms can dramatically reduce compute load and improve latency, particularly for common or predictable queries. When operated as public infrastructure, AIDNs ensure that AI services remain accessible even in environments with limited connectivity or bandwidth volatility.
Networking nodes collaborate to share state, KV caches, and partial computations across the network \cite{threadgill2024survey}.

Networking-layer intelligence becomes even more critical when inference is decomposed across the computing continuum. Contemporary orchestrators such as CODECO illustrate how real-time telemetry, including energy consumption at a node, set of nodes, and network (link, flow) can guide the placement of AI workloads. In this model, simple tasks may remain on the edge, data-sensitive operations may move to nearby micro–data centers, and only the most demanding reasoning steps escalate to the cloud. This form of context-aware migration turns the network into a mechanism of fairness: it ensures that limited bandwidth does not become a gatekeeper to high-quality AI services and that communities with constrained connectivity are not relegated to inferior AI performance.

The data infrastructure underpinning AI also requires re-thinking. Retrieval-augmented systems depend on large vector databases and indexing layers that are presently controlled by private cloud operators. For AI to serve as a social infrastructure, these retrieval layers must themselves become public resources: replicated across regional nodes, transparently governed, and accessible without discriminatory pricing. Locality-aware data storage reduces energy consumption by limiting unnecessary data transport, improves resilience by reducing dependence on centralized providers, and directly supports equitable access to knowledge-intensive AI capabilities.

\subsection{Public AI Utilities and Inference Commons}
Treating AI as public social infrastructure ultimately requires operational models modeled on existing public utilities. An example can be derived from the smart grid models that govern modern electricity distribution \cite{amin2024smart}. In a smart grid, demand is balanced with supply through real-time monitoring, differentiated pricing, distributed generation, and fairness-oriented allocation strategies. These mechanisms ensure that electricity remains universally accessible even under scarcity, while preventing greedy consumption that could jeopardize long-term sustainability. A similar logic must guide the creation of public AI utilities, where inference and reasoning capabilities are managed not as unlimited commodities but as shared societal resources subject to responsible governance.

Like electricity, AI inference is intrinsically resource-intensive: it demands energy, network bandwidth, and compute cycles. When market incentives dominate allocation, access is often rationed through subscription tiers, latency prioritization, or scarcity pricing—patterns that mirror peak-demand dynamics in unmanaged energy grids \cite{el2023evaluating}. Such mechanisms are incompatible with the idea of access to AI as an intergenerational civil right. Public AI utilities therefore require resource budgeting, priority allocation, transparent accounting, and adaptive routing, functions that closely mirror the governance principles underpinning modern smart-grid systems.

Understanding AI as a digital public good helps clarify this duality. Although AI-generated knowledge value does not diminish with use, the production of that information depends on scarce computational and energy resources. Scholars of the knowledge commons have long emphasized that such informational goods still require public oversight to prevent exclusion and ensure equitable access \cite{bollier2025think}. Public AI utilities extend this logic: they preserve broad access to AI’s non-depletable informational benefits while managing the competing compute and energy infrastructure necessary to produce them. 

Looking more long-term, such public good perspective would also align with the limited lifetime for Intellectual Property (IP) rights protection, eventually transferring any protected knowledge, including fundamental AI models derived from ingested data, into the public domain. Balancing commercial and public interests here demand that public access to that (ultimately) public knowledge is ensured for generations to come, thus not being purely commercially driven when it comes to accessing and utilising it. This already applies to those models that have been trained on already open knowledge, thus works that no longer underlie any copyright or patent protection. 

This governance role is carried out by a public orchestration layer, conceptually analogous to a grid operator in the energy sector. An analogy can also be drawn based on the current orchestration of applications across IoT-Edge-Cloud. Drawing on cluster management systems such as Kubernetes\footnote{https://kubernetes.io/} which has been shown to enable fair resource sharing and, based on extensions, to be able to support multi-tenant application workload management, the orchestration layer allocates inference workloads across the IoT-Edge-Cloud continuum in accordance with assigned priorities. Unlike commercial operators that optimize for throughput or profit, a public orchestrator can enforce equitable access, energy ceilings, and fairness constraints, particularly during periods of resource scarcity. These decisions are not purely technical; they represent a form of democratic governance over critical digital infrastructure \cite{plantin2018infrastructure}).

Within this system, AI capabilities are delivered through an open \textit{inference commons}, i.e., a shared pool of models, retrieval layers, and computational resources accessible to all citizens and public institutions. Theoretical foundations for commons and infrastructure governance \cite{bollier2025think, tilson2010research} support the idea that complex, interdependent systems can be managed sustainably and equitably when treated as shared societal assets. Federated and edge-based deployment models \cite{satyanarayanan2017emergence, arroba2024sustainable} ensure that computation is executed as close to users as possible. Local micro–data centers, analogous to substations in electrical grids, can provide public institutions such as schools, hospitals, administrative agencies with stable access to inference resources while reducing reliance on distant cloud providers. They also enforce data locality principles, thereby limiting the environmental costs of long-distance data movement and addressing well-documented concerns around centralized data extraction and power imbalances \cite{crawford2021atlas}.

The analogy to smart grids further clarifies the role of pricing and demand-shaping mechanisms in a public AI utility. Rather than monetizing outputs, public AI systems can employ usage-aware scheduling and energy-indexed inference budgets to stabilize demand during peak usage periods. Such mechanisms reflect insights from economic and regulatory models of utility pricing and prevent inequities in which high-income users receive privileged access while others face degraded service.

\section{A 3D Architecture for AI Delivery Networks Supporting Intergenerational Civil Rights}\label{sec:strawman}
\label{3D-Arch}
This section outlines an architecture for delivering AI at global scale. For this, we position our key decentralization principles driving our design along the technological trajectories we discussed in Section \ref{sec:trajectories}, followed by a strawman architecture for AIDNs, and a public good driven deployment sketch.

\subsection{3D Principles}
Decentralization plays a key role in the technological attempt to provide the balance between energy and access, captured in the ICR. In our view, decentralization needs to be provided along three key dimensions:

\textbf{Decentralize to contextualize:} While much of today’s AI discourse is driven by centrally trained large language models (LLMs), we argue that locally relevant data will be critical to the next generation of AI services. This relevance extends beyond data ingestion for fine-tuning general-purpose models to specific use cases, geographies, and populations; it also applies to the contextualization of inference over locally grounded data. Many application domains—including agriculture, manufacturing, healthcare, transportation, and others—will increasingly depend on such contextualization.

We contend that decentralizing the capabilities required to contextualize AI models and to perform inference is essential to ensure data freshness, compliance with local privacy regulations, and relevance of results. The resulting localization of computation also reduces latency by serving queries from AI nodes closer to end users, rather than relying solely on traffic engineering in high-tier networks. This approach supports more equitable access to contextualized knowledge among users.

From a networking perspective, localization further alleviates congestion in the core network, allowing scarce backbone resources to be reserved for ultra-large data transfers that distribute foundational AI models to locations where they can be contextualized for local use.

\textbf{Decentralize to rebalance energy demands:} In order to counter the adverse effects of centralized AI model training in ultra-large data centres, decentralization, together with the contextualization of models and use, will allow for rebalancing energy demands from central points of ultra-high energy consumption towards edge energy generation and consumption, where computing for localized finetuning of models and contextual inferencing can be balanced with centrally generated foundational knowledge. Advances in SW, specifically virtualization platforms, such as Kubernetes and others, can be employed to seamlessly transfer computing where it is needed and where it is most efficiently executed, even optimizing the use of surplus energy in locations of the sustainable energy producing industry, thus driving the ambition for a truly carbon-neutral (or even negative) AI delivery fabric, particularly through including carbon-neutral policies in the deployment of our architecture. This could be spearheaded, for instance, through public players, while also creating a competitive local market for providing surplus energy to direct adaptive AI traffic to those computational entities that use surplus energy most effectively. This will require advances in collective communication semantics that incorporate energy metrics as a key input into the traffic steering decision.  

\textbf{Decentralize to scale:} A balance between the skyrocketing demand for AI services, discussed in Section II, and a suitable supply of AI infrastructure cannot be fulfilled by centralized structures – the lessons learned from the global scale Internet development demonstrate this every day, most prominently with the overlay utilized for providing us all with affordable and massively scaled video content through CDNs. Its decentralized nature leads the way to devise similar structures for delivering AI services, albeit with crucial differences in balancing energy demands and contextualizing AI knowledge, as expressed in the previous two decentralization principles. As the flip side to global scale, AI delivery methods will need deep integration into the many subsystems driving decentralized, local delivery to industries and end users, such as but not limited to future 6G systems.

\subsection{Strawman Architecture for AIDN}

\begin{figure*}[t]
  \centering
  \includegraphics[width=\textwidth]{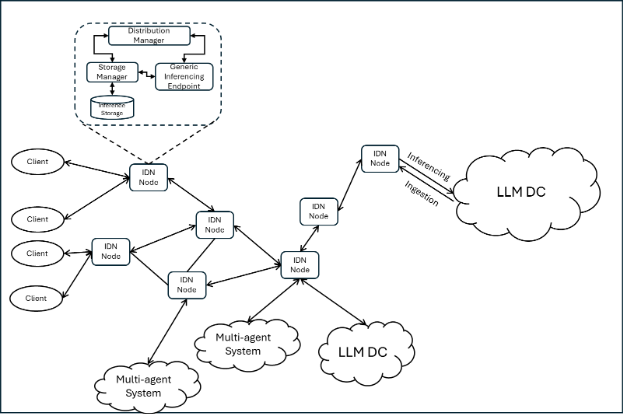}
  \caption{Strawman architecture for a global-scale AIDN.}
  \label{fig:strawman}
\end{figure*}

Figure \ref{fig:strawman} outlines a strawman architecture for a global-scale realization of our 3D principles as a multi-domain overlay, where decentralized AIDN nodes are collaboratively pushing AI model and inferencing knowledge closer to the (AI) end user. As discussed in Section \ref{sec:trajectories}, this collaborative caching resembles that utilized in CDNs today for content distribution, albeit with marked differences in the functionalities of the participating nodes, as we discuss in the following. 

The basic knowledge building block in our strawman is a KV cache entry, generated at the origin node and injected into the AIDN overlay\footnote{KV caches here denote merely an example of a basic representation of a knowledge building block that may be utilized for distributed inferencing and may be replaced by suitable other solutions that will be developed in ongoing AI research and developments.}. Such entry is similar to a chunk in a video file, which combined with other KV cache entries provides the contextualized knowledge over which to infer. KV cache entries are combined to larger KV caches, which are streamed across the AIDN network to serve as localized knowledge to be used in inferencing request arriving from the end user side. 

Each participating AIDN node is comprised of three key elements: 

\begin{itemize}
    \item The \textit{Storage Manager (SM)} manages the received KV cache entries to build local AI knowledge, enriching the KV caches with suitable metadata to respond to received inferencing requests but also to refine received KV caches with locally ingested, e.g., IoT data. 
    \item The \textit{Distribution Manager (DM)} is responsible to push KV caches further downstream in the global overlay while also actively pulling KV cache entries from nodes further upstream, ultimately from the origin of the AI knowledge. Contextual data on historical usage, predicted usage, and locally ingested data is used to drive the push/pull strategies of the distribution manager. 
    \item Complementing the storage focused elements, the \textit{Generic Inferencing Endpoint (GIE)} provides the endpoint towards other AIDN nodes, ultimately the end users themselves at the inferencing level. Here, incoming inferencing requests drive local AI knowledge retrieval strategies as well as distribution strategies from and to other AIDN nodes. But unlike CDNs for today's video distribution, the GIE also performs computational operations to generate suitable responses to incoming inferencing requests, if the local AI knowledge exists or could be provided in suitable time from other AIDN nodes. If neither is possible, the GIE will forward the inferencing requests further upstream, ultimately to the origin of AI knowledge.
\end{itemize}

Although AIDNs share certain characteristics with CDNs and the streaming paradigms used in video distribution, we expect that existing protocols such as HTTP, QUIC, and related mechanisms will require adaptation to support the contextual nature of AI knowledge and the decentralization needed to rebalance energy demands. Guided by these decentralization principles, metadata at both the knowledge and energy levels will play a significantly expanded role. This metadata will not only annotate key–value (KV) cache content, but also actively inform protocol behavior, including the push–pull strategies of the DM and workload distribution mechanisms that determine the placement of computational capabilities within GIEs for localized inference.

\subsection{A Public Good Deployment}

he AIDN architecture, grounded in its three-dimensional principles, aligns closely with the discussion of AI as a public utility presented in Section \ref{sec:trajectories}. An instructive analogy is the global network of public libraries, which combines decentralized access to written knowledge with locally provided services that make this knowledge relevant and useful to local communities. In this sense, each library functions as a real-world analogue of an AIDN node within a decentralized network.

Similar public-good incentives to those that establish and sustain networks of public libraries are required to underpin the proposed 3D-based AIDN architecture, in order to achieve an appropriate balance between broad access and energy sustainability. Such incentives must extend beyond the predominantly commercial nature of today’s CNDNs, in which a small number of dominant actors effectively control content distribution, delivery objectives, and performance, often through tiered pricing models that regulate access.

A driver for a more public good oriented incentive structure, we believe, may stem from the data that flows into the AIDN  architecture, where the public interest here is driven by the policies for ingestion of data, the contextualization of inferencing, and the protection afforded to business entities and citizen alike within such digital environment, such as GDPR and the EU AI act.

Such public interest may drive a model in which public interest is not embodied in the policies governing the digital environment but where parts of the infrastructure itself is being positioned as a public good, invested in and operated by a public private partnership (PPP) that places public interests alongside the commercial ones for profit maximization. 

The inherently decentralized AIDN architecture provides a strong foundation for establishing a combined public–private partnership capable of supporting a truly global AI delivery infrastructure.

\section{Conclusions}
\label{conclusions}

This paper has argued that the future of AI cannot be governed solely through the lenses of performance, innovation, or market efficiency. AI systems have become deeply embedded in education, employment, administration, and access to knowledge, and are increasingly indispensable for meaningful participation in social, economic, and political life. At the same time, their deployment relies on finite and unevenly distributed energy, networking, and computational resources. These dual characteristics place AI at the intersection of civil rights and intergenerational justice.

We showed that citizen access to AI, although often implicitly assumed to be abundant or universal, will inevitably face increasing constraints. Scaling demand for inference, persistent infrastructural bottlenecks, rising energy consumption, and existing digital inequalities all exert structural pressure toward restricted or tiered access. Yet current regulatory and governance frameworks largely overlook this tension. While they address safety, accountability, and discrimination, they fail to recognize citizens’ normative claims to AI systems derived from collectively produced public knowledge, and they do not provide mechanisms to balance equitable access with long-term resource sustainability.

To address this gap, we proposed recognizing access to AI as an \emph{Intergenerational Civil Right}. Framing AI access in this way binds its development to two non-negotiable obligations: ensuring equitable access for present-day citizens and safeguarding the environmental and infrastructural conditions required for future generations. This perspective reframes AI not as a discretionary commercial service, but as a form of shared social infrastructure, comparable to education, public knowledge systems, or essential communication networks.

We further argued that such a right cannot be upheld through abstract principles alone. It must be instantiated through concrete technical and architectural choices. To this end, we outlined technological trajectories consistent with AI as a public infrastructure, including decomposed and orchestrable AI workloads, energy-aware inference, modular and sustainable model architectures, and networking stacks designed for locality, caching, and fairness. Building on these principles, we sketched a strawman architecture for AIDNs that extends content distribution paradigms toward distributed, energy-conscious, and context-aware inference, suitable for governance as a public utility.

The trajectory of AI is not predetermined. Decisions made today about system architecture, governance, and access will shape whether AI entrenches inequality and environmental strain or becomes a durable public resource that enhances human capability across generations. Recognizing access to AI as an intergenerational civil right provides a unifying framework to align legal obligations, social values, and system design, ensuring that AI remains both accessible and sustainable. The challenge ahead is not merely to regulate AI, but to deliberately design and govern it as part of the shared infrastructural foundations of society.


\bibliographystyle{ACM-Reference-Format}
\bibliography{conference_101719}

@book{crawford2021atlas,
  title     = {The Atlas of {AI}: Power, Politics, and the Planetary Costs of Artificial Intelligence},
  author    = {Crawford, Kate},
  year      = {2021},
  publisher = {Yale University Press}
}

@article{arroba2024sustainable,
  title   = {Sustainable edge computing: Challenges and future directions},
  author  = {Arroba, Patricia and Buyya, Rajkumar and C{\'a}rdenas, Rom{\'a}n and Risco-Mart{\'\i}n, Jos{\'e} L and Moya, Jos{\'e} M},
  journal = {Software: Practice and Experience},
  volume  = {54},
  number  = {11},
  pages   = {2272--2296},
  year    = {2024},
  publisher = {Wiley Online Library}
}

@article{satyanarayanan2017emergence,
  title   = {The emergence of edge computing},
  author  = {Satyanarayanan, Mahadev},
  journal = {Computer},
  volume  = {50},
  number  = {1},
  pages   = {30--39},
  year    = {2017},
  publisher = {IEEE}
}

@article{tilson2010research,
  title   = {Research commentary---Digital infrastructures: The missing {IS} research agenda},
  author  = {Tilson, David and Lyytinen, Kalle and S{\o}rensen, Carsten},
  journal = {Information Systems Research},
  volume  = {21},
  number  = {4},
  pages   = {748--759},
  year    = {2010},
  publisher = {INFORMS}
}

@article{plantin2018infrastructure,
  title   = {Infrastructure studies meet platform studies in the age of {Google} and {Facebook}},
  author  = {Plantin, Jean-Christophe and Lagoze, Carl and Edwards, Paul N and Sandvig, Christian},
  journal = {New Media \& Society},
  volume  = {20},
  number  = {1},
  pages   = {293--310},
  year    = {2018},
  publisher = {SAGE Publications}
}

@book{bollier2025think,
  title     = {Think Like a Commoner: A Short Introduction to the Life of the Commons},
  author    = {Bollier, David},
  year      = {2014},
  publisher = {New Society Publishers}
}

@article{el2023evaluating,
  title   = {Evaluating demand charges as instruments for managing peak-demand},
  author  = {El Gohary, Fouad and Stikvoort, Britt and Bartusch, Cajsa},
  journal = {Renewable and Sustainable Energy Reviews},
  volume  = {188},
  pages   = {113876},
  year    = {2023},
  publisher = {Elsevier}
}

@article{amin2024smart,
  title   = {Smart Grids: A Comprehensive Review of Technologies, Challenges, and Future Directions},
  author  = {Amin, Rameez and Kazmi, Abdullah Ali},
  journal = {Journal of Engineering and Computational Intelligence Review},
  volume  = {2},
  number  = {2},
  pages   = {45--70},
  year    = {2024}
}

@article{threadgill2024survey,
  title   = {A Survey of Distributed Learning in Cloud, Mobile, and Edge Settings},
  author  = {Threadgill, Madison and Gerstlauer, Andreas},
  journal = {arXiv preprint arXiv:2405.15079},
  year    = {2024}
}

@article{oche2025systematic,
  title   = {A systematic review of key retrieval-augmented generation ({RAG}) systems: Progress, gaps, and future directions},
  author  = {Oche, Agada Joseph and Folashade, Ademola Glory and Ghosal, Tirthankar and Biswas, Arpan},
  journal = {arXiv preprint arXiv:2507.18910},
  year    = {2025}
}

@inproceedings{li2024llm,
  title     = {{LLM} inference serving: Survey of recent advances and opportunities},
  author    = {Li, Baolin and Jiang, Yankai and Gadepally, Vijay and Tiwari, Devesh},
  booktitle = {2024 IEEE High Performance Extreme Computing Conference (HPEC)},
  pages     = {1--8},
  year      = {2024},
  organization = {IEEE}
}

@article{10630483,
  author   = {Kristiansen N{\o}land, Jonas and Hjelmeland, Martin and Korp{\aa}s, Magnus},
  journal  = {IEEE Access},
  title    = {Will Energy-Hungry {AI} Create a Baseload Power Demand Boom?},
  year     = {2024},
  volume   = {12},
  number   = {},
  pages    = {110353--110360},
  doi      = {10.1109/ACCESS.2024.3440217}
}

@article{rahmath2024early,
  title   = {Early-exit deep neural network---a comprehensive survey},
  author  = {Rahmath P, Haseena and Srivastava, Vishal and Chaurasia, Kuldeep and Pacheco, Roberto G and Couto, Rodrigo S},
  journal = {ACM Computing Surveys},
  volume  = {57},
  number  = {3},
  pages   = {1--37},
  year    = {2024},
  publisher = {ACM}
}

@article{mu2025comprehensive,
  title   = {A comprehensive survey of mixture-of-experts: Algorithms, theory, and applications},
  author  = {Mu, Siyuan and Lin, Sen},
  journal = {arXiv preprint arXiv:2503.07137},
  year    = {2025}
}

@article{gao2024modular,
  title   = {Modular {RAG}: Transforming {RAG} systems into lego-like reconfigurable frameworks},
  author  = {Gao, Yunfan and Xiong, Yun and Wang, Meng and Wang, Haofen},
  journal = {arXiv preprint arXiv:2407.21059},
  year    = {2024}
}

@inproceedings{hu2022lora,
  title     = {{LoRA}: Low-rank adaptation of large language models},
  author    = {Hu, Edward J and Shen, Yelong and Wallis, Phillip and Allen-Zhu, Zeyuan and Li, Yuanzhi and Wang, Shean and Wang, Lu and Chen, Weizhu and others},
  booktitle = {International Conference on Learning Representations ({ICLR})},
  year      = {2022}
}

@article{han2015deep,
  title   = {Deep compression: Compressing deep neural networks with pruning, trained quantization and {Huffman} coding},
  author  = {Han, Song and Mao, Huizi and Dally, William J},
  journal = {arXiv preprint arXiv:1510.00149},
  year    = {2015}
}

@article{patterson2021carbon,
  title   = {Carbon emissions and large neural network training},
  author  = {Patterson, David and Gonzalez, Joseph and Le, Quoc and Liang, Chen and Munguia, Lluis-Miquel and Rothchild, Daniel and So, David and Texier, Maud and Dean, Jeff},
  journal = {arXiv preprint arXiv:2104.10350},
  year    = {2021}
}

@inproceedings{strubell2019energy,
  title     = {Energy and policy considerations for deep learning in {NLP}},
  author    = {Strubell, Emma and Ganesh, Ananya and McCallum, Andrew},
  booktitle = {Proceedings of the 57th Annual Meeting of the Association for Computational Linguistics},
  pages     = {3645--3650},
  year      = {2019}
}

@article{wu2022memorizing,
  title   = {Memorizing transformers},
  author  = {Wu, Yuhuai and Rabe, Markus N and Hutchins, DeLesley and Szegedy, Christian},
  journal = {arXiv preprint arXiv:2203.08913},
  year    = {2022}
}

@article{peng2023rwkv,
  title   = {{RWKV}: Reinventing {RNNs} for the transformer era},
  author  = {Peng, Bo and Alcaide, Eric and Anthony, Quentin and Albalak, Alon and Arcadinho, Samuel and Biderman, Stella and Cao, Huanqi and Cheng, Xin and Chung, Michael and Grella, Matteo and others},
  journal = {arXiv preprint arXiv:2305.13048},
  year    = {2023}
}

@article{an2024make,
  title   = {Make your {LLM} fully utilize the context},
  author  = {An, Shengnan and Ma, Zexiong and Lin, Zeqi and Zheng, Nanning and Lou, Jian-Guang and Chen, Weizhu},
  journal = {Advances in Neural Information Processing Systems},
  volume  = {37},
  pages   = {62160--62188},
  year    = {2024}
}

@article{shi2024keep,
  title   = {Keep the cost down: A review on methods to optimize {LLM}'s {KV}-cache consumption},
  author  = {Shi, Luohe and Zhang, Hongyi and Yao, Yao and Li, Zuchao and Zhao, Hai},
  journal = {arXiv preprint arXiv:2407.18003},
  year    = {2024}
}

@article{dao2022flashattention,
  title   = {{FlashAttention}: Fast and memory-efficient exact attention with {IO}-awareness},
  author  = {Dao, Tri and Fu, Dan and Ermon, Stefano and Rudra, Atri and R{\'e}, Christopher},
  journal = {Advances in Neural Information Processing Systems},
  volume  = {35},
  pages   = {16344--16359},
  year    = {2022}
}

@incollection{stiglitz1999knowledge,
  title     = {Knowledge as a Global Public Good},
  author    = {Stiglitz, Joseph E},
  booktitle = {Global Public Goods: International Cooperation in the 21st Century},
  editor    = {Kaul, Inge and Grunberg, Isabelle and Stern, Marc A},
  pages     = {308--325},
  publisher = {Oxford University Press},
  address   = {New York},
  year      = {1999}
}

@article{riva2025invisible,
  title   = {Invisible Architectures of Thought: Toward a New Science of {AI} as Cognitive Infrastructure},
  author  = {Riva, Giuseppe},
  journal = {arXiv preprint arXiv:2507.22893},
  year    = {2025}
}

@inproceedings{languages,
  title     = {{C}hat{GPT} Beyond {E}nglish: Towards a Comprehensive Evaluation of Large Language Models in Multilingual Learning},
  author    = {Lai, Viet Dac and Ngo, Nghia and Pouran Ben Veyseh, Amir and Man, Hieu and Dernoncourt, Franck and Bui, Trung and Nguyen, Thien Huu},
  editor    = {Bouamor, Houda and Pino, Juan and Bali, Kalika},
  booktitle = {Findings of the Association for Computational Linguistics: EMNLP 2023},
  month     = dec,
  year      = {2023},
  address   = {Singapore},
  publisher = {Association for Computational Linguistics},
  url       = {https://aclanthology.org/2023.findings-emnlp.878/},
  doi       = {10.18653/v1/2023.findings-emnlp.878},
  pages     = {13171--13189}
}

@article{vspecian2025universal,
  title   = {Universal Basic {AI} Access: Countering the Digital Divide},
  author  = {{\v{S}}peci{\'a}n, Petr and {\v{S}}peci{\'a}nov{\'a}, Jitka and others},
  journal = {Acta Informatica Pragensia},
  volume  = {14},
  number  = {2},
  pages   = {272--281},
  year    = {2025},
  publisher = {Acta Informatica Pragensia}
}

@article{bo2025oecd,
  title   = {{OECD} digital education outlook 2023: Towards an effective education ecosystem},
  author  = {Bo, Nang Sagawah Win},
  journal = {Hungarian Educational Research Journal},
  volume  = {15},
  number  = {2},
  pages   = {284--289},
  year    = {2025},
  publisher = {Akad{\'e}miai Kiad{\'o} Budapest}
}

@article{sofia2024framework,
  author  = {Rute, Sofia and others},
  title   = {A Framework for Cognitive, Decentralized Container Orchestration},
  journal = {IEEE Access},
  year    = {2024},
  volume  = {12},
  pages   = {79978--80008},
  doi     = {10.1109/ACCESS.2024.3406861}
}

@article{InformationDemocracy2024,
  title  = {{AI} as a Public Good: Ensuring Democratic Control of {AI}},
  author = {{Forum on Information and Democracy}},
  note   = {White paper / policy report},
  url    = {https://informationdemocracy.org/wp-content/uploads/2024/03/ID-AI-as-a-Public-Good-Feb-2024.pdf},
  year   = {2024}
}

@online{UNESCO2021_EthicsAI,
  title   = {Recommendation on the Ethics of Artificial Intelligence},
  author  = {{UNESCO}},
  year    = {2021},
  url     = {https://unesdoc.unesco.org/ark:/48223/pf0000381137},
  urldate = {2025-03-09}
}

@article{Freilich2025_DataAsPolicy,
  title   = {Data as Policy},
  author  = {Freilich, Janet and Price, W. Nicholson II},
  year    = {2026},
  journal = {Boston College Law Review},
  note    = {Forthcoming},
  url     = {https://scholarship.law.bu.edu/cgi/viewcontent.cgi?article=5011&context=faculty_scholarship}
}

@misc{NIST_AIRMF2023,
  title        = {Artificial Intelligence Risk Management Framework ({AI} {RMF} 1.0)},
  author       = {{National Institute of Standards and Technology}},
  year         = {2023},
  howpublished = {\url{https://www.nist.gov/itl/ai-risk-management-framework}}
}

@misc{USEO2023,
  title        = {Executive Order on the Safe, Secure, and Trustworthy Development and Use of Artificial Intelligence},
  author       = {{The White House}},
  year         = {2023},
  howpublished = {\url{https://www.whitehouse.gov/briefing-room/presidential-actions/}},
  note         = {Executive Order, October 30, 2023}
}

@misc{CRS_USAI,
  title        = {Artificial Intelligence and National Policy},
  author       = {{Congressional Research Service}},
  year         = {2023},
  howpublished = {\url{https://crsreports.congress.gov}}
}

@misc{State_AI_Overview,
  title        = {Overview of US State-Level {AI} Legislation},
  author       = {{National Conference of State Legislatures}},
  year         = {2024},
  howpublished = {\url{https://www.ncsl.org/technology-and-communication/artificial-intelligence}}
}

@misc{WeylLanier_PublicOptionAI,
  author = {Weyl, Glen and Lanier, Jaron},
  title  = {The Public Option for {AI}},
  year   = {2023},
  url    = {https://www.radicalxchange.org/media/blog/public-option-for-ai}
}

@article{Vinuesa2019_AIforSDGs,
  title   = {The Role of Artificial Intelligence in Achieving the Sustainable Development Goals},
  author  = {Vinuesa, Ricardo and Azizpour, Hossein and Leite, Iolanda and Balaam, Madeleine and Dignum, Virginia and Domisch, Sami and Fell{\"a}nder, Anna and Langhans, Simone and Tegmark, Max and Fuso Nerini, Francesco},
  journal = {Nature Communications},
  volume  = {11},
  number  = {1},
  pages   = {233},
  year    = {2020},
  doi     = {10.1038/s41467-019-14108-9}
}

@misc{UN_UDHR_1948,
  author = {{United Nations}},
  title  = {Universal Declaration of Human Rights},
  year   = {1948},
  url    = {https://www.un.org/en/about-us/universal-declaration-of-human-rights}
}

@online{IEA2025_EnergyAndAI,
  title   = {Energy and {AI}},
  author  = {{International Energy Agency}},
  year    = {2025},
  url     = {https://www.iea.org/reports/energy-and-ai},
  urldate = {2025-03-09},
  license = {CC BY 4.0}
}

@misc{EU_AI_Act_2024,
  title        = {Artificial Intelligence Act},
  author       = {{European Union}},
  howpublished = {Regulation (EU) 2024/1689 of the European Parliament and of the Council of 13 June 2024 laying down harmonised rules on artificial intelligence},
  year         = {2024},
  note         = {Official Journal of the European Union, L 2024/1689 (12 July 2024). Entered into force on 1 August 2024},
  url          = {https://eur-lex.europa.eu/eli/reg/2024/1689/oj}
}

@misc{UN2023_GoverningAI,
  title        = {Governing {AI} for Humanity: Interim Report},
  author       = {{United Nations High-Level Advisory Body on Artificial Intelligence}},
  year         = {2023},
  howpublished = {United Nations},
  url          = {https://www.un.org/en/ai-advisory-body/interim-report}
}

@article{Kaplan2020ScalingLaws,
  title   = {Scaling Laws for Neural Language Models},
  author  = {Kaplan, Jared and others},
  journal = {arXiv preprint arXiv:2001.08361},
  year    = {2020}
}

@book{Eubanks2018AutomatingInequality,
  author    = {Eubanks, Virginia},
  title     = {Automating Inequality: How High-Tech Tools Profile, Police, and Punish the Poor},
  publisher = {St. Martin's Press},
  year      = {2018},
  address   = {New York},
  isbn      = {978-1-250-07908-5}
}

@online{IamIP2025WhyPatentsExpire,
  author = {IamIP},
  title  = {Why Patents Expire and What Does That Mean for Businesses},
  year   = {2025},
  url    = {https://iamip.com/why-patents-expire-and-what-does-that-mean-for-businesses/}
}

@article{Price2017ExpiredPatents,
  author  = {Price, W. Nicholson},
  title   = {Expired Patents, Trade Secrets, and Stymied Competition},
  journal = {Notre Dame Law Review},
  volume  = {92},
  number  = {4},
  pages   = {1611--1640},
  year    = {2017}
}

@online{AIStackExchangeEnergy,
  title   = {How much energy consumption is involved in ChatGPT responses being generated?},
  author  = {{AI Stack Exchange}},
  year    = {2023},
  url     = {https://ai.stackexchange.com/questions/38970/how-much-energy-consumption-is-involved-in-chat-gpt-responses-being-generated}
}
\end{document}